# Understanding Barriers to Internal Startups in Large Organizations: Evidence from a Globally Distributed Company

Tor Sporsem
*SINTEF*
Trondheim, Norway
tor.sporsem@sintef.no

Anastasiia Tkalich
*SINTEF*
Trondheim, Norway
anastasiia.tkalich@sintef.no

Nils Brede Moe
*SINTEF*
Trondheim, Norway
nils.b.mow@sintef.no

Marius Mikalsen
*SINTEF*
Trondheim, Norway
marius.mikalsen@sintef.no

*Abstract*—Large global companies need to speed up their innovation activities to increase competitive advantage. However, such companies' organizational structures impede their ability to capture trends they are well aware of due to bureaucracy, slow decision-making, distributed departments, and distributed processes. One way to strengthen the innovation capability is through fostering internal startups. We report findings from an embedded multiple-case study of five internal startups in a globally distributed company to identify barriers for software product innovation: *late involvement of software developers*, *executive sponsor is missing or not clarified*, *yearly budgeting and planning*, *unclear decision-making authority*, *lack of digital infrastructure for experimentation* and *access to data from external actors*. Drawing on the framework of continuous software engineering proposed by Fitzgerald and Stol, we discuss the role of BizDev in software product innovation. We suggest that lack of continuity, rather than the lack of speed, is an ultimate challenge for internal startups in large global companies.

## I. INTRODUCTION

Large, distributed software-intensive companies must innovate at high speed to maintain their position in the changing global market and compete with an increasing number of startups. According to the Society for Information Management report, innovation is the 4[th] major IT management issue [1]. While global software engineering (GSE) is a common component of most large companies, a considerable share of global projects still does not meet the expectations regarding time to market [2]. To address the challenge of bringing new products to the market faster, many large companies tend to engage in a systematic way of innovation [3]. One suggested tool for this is corporate innovation (corporate entrepreneurship) [4]. Lean startup is a prominent example of corporate entrepreneurship approach, which is argued to improve large companies' innovativeness [5]. Eric Ries formulated the Lean startup principles to create new businesses through early feedback from potential customers and iterative product building [6]. It has been argued that corporate entrepreneurship in general and Lean startup in particular can increase firms' ability to innovate [4], [6].

Even though globally distributed companies have a high focus on new product development, the development process is not without barriers. The low availability of crucial experts and the absence of organizational support for effective coordination across sites reduce development speed [7]. Further, today's world with a constantly changing market and organizational structure requires flexibility. However, earlier research shows that startups' development process in large companies can be hindered due to bureaucracy [8]. Barriers can be several layers of approval, lack of necessary autonomy for innovators [3], interference with the existing business model [9], and low fit with standard procedures in the organization [5]. Further, there is often a need to support the customer in the uptake of new products (sensing customer behavior and adapting the digital solutions), which require a change in the internal organization [10]. Such barriers reduce innovation speed and thus jeopardize its very purpose – to acquire profit from the market where no competition exists until the competitors introduce similar products [11]. One suggested way to improve software product development speed is a continuous integration between business and development coined *BizDev* [12]. There has not been sufficient empirical data that contributes to the understanding of the concept and its value for software product innovation, while integration between development and business is now very high on the IT-management agenda [1].

Large companies need to deliver on their existing products and are therefore less focused on the ideas beyond their scope [13]. Accordingly, GSE companies need to have a holistic view of their innovation processes and practices. Given the complex landscape in which global software projects operate, we need to understand how they apply innovation processes and what the barriers are for the innovation speed. Thus, this study aims to answer the following research question:

*What are the barriers to software product innovation speed in a large company?*

To answer, we report an empirical insight from a case study. We draw on the framework of continuous software engineering suggested by Fitzgerald and Stol [12] to make sense of our evidence. We chose this framework because it describes the link between software development and business strategy (BizDev).

The remainder of the paper is organized as follows. Chapter II describes related work and the conceptual framework. Chapter III outlines our research method, followed by chapter IV that presents our findings. In chapter V we discuss the results through the lens of the conceptual framework of continuous software engineering and show implications for practice. Chapter VI conclude our work and point to future research.

## II. BACKGROUND

This chapter offers an overview of software product innovation literature in large global companies (chapter II A) and presents a theoretical framework on continuous software engineering (chapter II B). Software product innovation in large global companies requires strategies to maintain continuous innovation processes and achieve high innovation speed. It is argued that to achieve this; there is a need for short development cycles to provide continuous and rapid loops of iterative learning, introduce improvements, and test them [14]. To gain speed, the software development processes must be continuously integrated across business, development, and operations [12].

### A. Software product innovation in large global companies and innovation speed

*Software product innovation* is defined as creating and introducing a novel software product to the market [15]. Startups within existing companies are known as *corporate innovation* or *corporate entrepreneurship* [3]. These are often risk-taking and proactive undertakings that can lead either to incremental (new product in the current market or new market for a current product) or radical innovation (a new product for new market) [3]. Innovation is something that comes naturally for a startup. However, it becomes harder in large global companies, and therefore, such companies need strategies to sustain the development of new products and features [16]. Google relies on processes like design sprints [17], and Atlassian relies on methods like "FedEx Day" where developers are given one day to showcase a proof of concept they believe should be part of the product. Atlassian also applies the method "20% Time" to allow more ambitious innovation projects to be undertaken [16]. A similar strategy was adopted much earlier at 3M, where the "15-percent rule" gives technical staff six hours per week on projects of their choosing. This strategy led to ScotchTape and Post-it Notes. *Lean startup* is another popular approach to corporate innovation when software is developed and validated through continuous experiments with stakeholders to minimize development costs and increase customer satisfaction [6]. The experimentation allows increasing the product's fit to market through continuous user's feedback and incremental improvements. It is argued that Lean startup increases the speed of product development [5] and product-market fit [3]. Lean startup is described through the following five principles [6]:

1. *Entrepreneurs are everywhere.* The startup mindset of startups should not be limited to only small young companies but can be applied to organizations of all sizes.
2. *Entrepreneurship is management.* Developing new products and services requires its own management approach to cope with extreme uncertainty throughout the development process.
3. *Validated learning.* There is a need to run frequent experiments to validate the entrepreneurs' assumptions on the actual market needs to deal with uncertainty.
4. *Build-Measure-Learn.* Business ideas should be implemented as soon as possible to acquire customer feedback, evaluate it and incorporate to improve the product.
5. *Innovation accounting.* The innovators should track their startups' success by creating and monitoring metrics, milestones etc.

Lean Startup is often called hypothesis-driven development [18]. Relying on hypothesis-driven development requires a framework for conducting continuous experiments. Fagerholm et al. [14] found that an experimentation system (the 4$^{th}$ principle of Lean Startup) requires the ability to release minimum viable products or features, design and manage experiment plans, link experiment results with a product roadmap, and manage a flexible business strategy. Further, they found that a challenge for continuous experimentation (the 3$^{rd}$ principle) is proper and rapid design of experiments, that there must exist feedback loops, and that the organization needs the capability of analyzing data from the experiment. Feedback loops are essential for relevant information to be fed back from experiments into several parts of the organization. The organization needs a workforce with the ability to collect and analyze both qualitative and quantitative data and the ability to properly define decision criteria and act on data-driven decisions [14]. Continuous learning is a key activity when doing continuous experimentation (the 4$^{th}$ principle). Khanna et al. [18] argued that the amount of learning entrepreneurs achieve depends on user involvement and their existing knowledge about the market, industry, and technology. Little user involvement might lead to little knowledge gained from testing hypotheses.

Large companies can engage in corporate innovation to enter into new competition for them, thus avoiding focusing only on their own business strategy [3]. By timely launching innovative products, firms can increase their profitability because they enjoy a temporary monopoly position until the competitors introduce a similar product [11]. Since being the first inventor is an advantage, speed is crucial for organizations to succeed with innovation. *Innovation speed* usually refers to the time spent on the innovation activities between the initial concept definition and the product's introduction to the marketplace [19]. Speedy introduction means reduced time to market, which also allows achieving revenue early, thus reducing the company's costs for developing the product [6]. Even though speed is generally seen as crucial for innovation, it is pointed out that speed is not always the most important. The utility of the innovation speed can, for example, be reduced when the competitive pressure of a firm is low [19]. In software development, it is argued that continuity and flow are more important for product development process than the speed itself [12].

### B. Conceptual framework: Continuous software engineering

Inspired by Lean startup, Continuous software engineering [12] is a holistic framework that describes the entire software lifecycle through three main phases: Business Strategy, Development, and Operations. *Business Strategy* refers to the activity of business management (e.g., planning and budgeting). *Development* comprises the main software activities of analysis, design, coding and verification. The *Operation* describes the use of the software by the employees or customers. The authors emphasize the importance of continuous linkage between the three phases of software development to secure *flow*. The concept of flow originates from Lean Thinking and agile methods and refers to continuous value-creating activities. For example, an identified product feature is immediately designed, integrated, tested, and deployed. It stands in contrast to the 'batch-and-queue' mentality, where a sequence of discrete activities are grouped in batches of the product and then queued for the next process step [12].

Fitzgerald & Stol [12] argue that there has been a narrow view concerning the topic of flow limited to agile practices only. They suggest carrying the concepts of flow to other activities surrounding software development, e.g., business activities like innovation, experimentation, planning, and budgeting [12]. In line with the concept of flow the authors describe continuous integration between Development and Operations (DevOps) and between Development and Business Strategy (BizDev) as crucial for software product development. *DevOps* has been described as the linkage between software development and operations, which allows better knowledge sharing between two traditionally separated teams (e.g., more frequent feedback and increased understanding of the customer among the developers; better understanding of the technology among the operations) [20]. Like DevOps, Fitzgerald & Stol [12] describe the need for integration between Development and Business Strategy, which they coin *BizDev*. To achieve BizDev the authors suggest that strategic planning and execution should be performed in collaboration between stakeholders from business and software (*continuous planning*). Additionally, frequent budgeting should be prioritized ahead of episodic budgeting (e.g., annual), as it allows for better alignment between the needs of business and those of software (*continuous budgeting*). Finally, *continuous innovation* is seen by the authors as a foundation of the software lifecycle. Continuous innovation seeks to establish a sustainable software development process in response to the conditions of the evolving market, thus integrating Business, Development and Operations.

Recent empirical and conceptual contributions to the literature on software development flow show that an increase in flow enhances software development team's performance indicators [21], [22]. Simultaneously, continuous processes are challenging in large companies and even more challenging in global companies. Both socio-technical and organizational factors have been found to have a significant influence on the continuous processes [23]. Examples of continuity challenges are continuous access to customers [24] and expertise [25]. If essential recourses are not available, important decisions are delayed, and problems stay unsolved. Because experts are often in other departments or locations and often occupied, they are hard to access [7]. One mitigating strategy to access experts in large distributed organizations is to have a strong social network [25], [26]. Another strategy is to include key experts in the team. The main success factor for, e.g., a continuous testing process, is to build one team out of developers and testers [23], [25].

III. METHOD

This chapter first describes the case and its context before presenting our data collection method and how we analyzed it.

*A. Case context*

MarComp (name suppressed for anonymity) is a global provider of business-to-business services within energy, business, digital and maritime industries. They are 13 000 employees located in offices in more than a hundred countries. The initiative for software product innovation has been part of the company strategy to shift towards digital products and services. Our case context was the organization's primarily maritime division that provides certification of ships, with a revenue of about 850 million dollars in 2019. The division's customers are ship owners and shipping management companies whose vessels are inspected by MarComp in order for them to comply with international standards of safety and quality. MarComp is operating and competing globally and hence considers digital products as crucial for offering value to its worldwide customers conveniently. The global nature of MarComp's operations means that they need to master global software product development to stay competitive. A solution developed in Scandinavia or at any other location can have multiple users all over the world. The IT department consists of software developers in Scandinavia and testers in China. Both the Norwegian employees and the Chinese employees are sitting together in their respective countries, making it easy for them to communicate and coordinate work locally. The IT department had used agile since 2008 and followed a Scrum-based process, where a regular release is 2-4 months long.

In 2019, during pressure for increased digitalization of operations, MarComp implemented an innovation framework to increase their software product development ability, which is heavily influenced by Lean startup principles. The maritime division was chosen to test the framework, which invited employees to submit new product ideas. After an 8-week campaign considering different ideas, they were left with 5 pilots who started developing their ideas using the innovation framework. The pilots can be considered internal startups, led by the employee submitting the idea who became a Product manager and responsible for realizing the idea. The innovation framework is a form of a stage-gate model and consists of six phases: Customer Insight, Viability, Proof of Concept, Build, Scale, and Sustain. Our data consists of all stages except Scale, and Sustain because none of the internal startups had reached these states yet. The innovation framework requires answering a set of predefined questions to progress to a new phase, such as *how many customers have committed to the proof-of-concept testing?* and *what is the estimated ROI?* Product managers answer to a Venture Board, consisting of various line managers within the division, who evaluate the answers and approves or disapproves the startup's progression to the next phase.

All 5 startups developed products for customers distributed around the world. They were located in Scandinavia but frequently crossed organizational borders spanning between departments in Europe and Asia. The product managers worked part time on their internal startups, balancing their usual responsibilities within operations. We do not describe each startup in detail in this paper due to space restrictions.

## B. Case Study, Data Collection and Analysis

We chose a longitudinal case study design [13]. We have been following the case since 2019 and report here findings from 5 startups that followed the innovation framework in MarCorp. We kept an exploratory approach as we did not set out to test any specific theory or hypothesis [27]. Also, we investigated the phenomenon broadly by talking to several different stakeholders, seeking to understand the context in which the startups operated. The data reported here was collected from June 2020 until December 2020 (see Table I below). We collected data in 3 different ways.

First, we conducted 10 interviews with different stakeholders, eight with product managers, one with the Venture Board facilitator, and one with the head of innovation process. The interviews were recorded and transcribed into 61 pages of text.

Second, we did two kinds of observations; we observed community of practice meetings where product managers shared knowledge, coordinated their efforts to other organizational units, and expanded their individual skills by inviting experts to give talks. Two researchers took notes in these meetings. We also did participatory observation by facilitating workshops concerning how internal startups adopted the innovation framework. One researcher facilitated the workshops, and another researcher took notes to document. Summaries of the workshops were sent to the participants. Lastly, we participated in the planning of the communities of practice meetings by giving research-based advice on how to conduct them

Third, we used documents. We collected documentation on the innovation framework that was being implemented in MarComp and data on the organization. We also got access to reports from the venture board on the status of the startups and e-mails.

Data analysis was performed in several steps. First, textual data was entered into the qualitative data analysis tool NVivo. Two researchers coded the data inductively, and 150 codes were created. They coded independently, compared and discussed codes in iterations. The second step was to arrange the codes into 31 themes. These themes were analyzed by all authors and grouped into six barriers for innovation speed. In this step, we also started to connect to theory and interpreted the barriers in light of the continuous software development framework [12]. Third, we presented temporary findings to the case (on the barriers) to get their feedback and adjust our interpretations.

TABLE I. EMPIRICAL DATA COLLECTION AND ANALYSIS

| Data source | Location | Time | Participants | Data gathered |
|---|---|---|---|---|
| *Interviews and informal conversations* | Virtual | Sep.2020 – Des. 2020 | 10 Interviews (5 product managers, 1 head of the innovation process, 1 facilitator) | Interviews on the startups, innovation process, work processes, context, stakeholders |
| *Participant observation (facilitating)* | Virtual | Jun. 2020 | Workshop: 8 participants (5 product managers, 3 line managers) | Meeting notes describing the innovation process, work processes, context, stakeholders |
|  | Face-to-face | Sep. 2020 | Workshop: 17 participants (6 product managers, 9 line managers, 1 facilitator, 1 head of the innovation process) |  |
|  | Virtual | Sep. 2020 – Des. 2020 | 4 planning meetings of Community of Practice meetings: varying number (3-6) of participants (product managers, line managers, facilitator, head of the innovation process) |  |
| *Observation of meetings* | Virtual | Jun. 2020 – Nov. 2020 | 4 Venture Board meetings: 8-17 participants (Venture Board members, product managers, head of the innovation process) |  |
|  | Virtual | Nov.2020 – Des. 2020 | 4 Community of Practice meeting: 8-26 participants (5-9 product managers, 2-10 line managers, 1 innovation framework coach, 1 head of the innovation process, 1 facilitator, 2 finance managers, 2 sales managers) |  |
|  | Virtual | Nov. 2020 | Knowledge sharing meeting: 13 participants (6 product managers, 5 line managers, 2 senior software developers) |  |
| *Documentation* | Virtual | Jun 2020 – Des 2020 | Descriptions of the organization, the innovation framework, and e-mail correspondence | Venture Board reports on status of startups, e-mails describing barriers to the innovation process, strategic documents |

## IV. RESULTS

To answer our research question – *what are the barriers to software product innovation speed in a large company* – we will now describe the 6 barriers to software product innovation that we identified in the case company. All barriers connected to the respective startups are listed in table II.

### A. Late involvement of software developers

Neither developers in Scandinavia nor testers in China entered the innovation process before the Build phase (see Case context for description of the phases). The reason was the necessity to validate business ideas with customers in the global market without investing too many resources upfront. The fear was that involving IT resources early would significantly increase cost and shift focus from evaluating the business concept to a technology project. From earlier projects, the experience was that a too strong focus on the technical aspects in the exploration phase led to products that the customer did not want. Further, involving IT resources from another department and even from another geographical site was challenging as it needed reprioritization of the already booked resources. However, this approach made the early stages of the innovation process disconnected from technology and led to two issues.

TABLE II. OBSERVED BARRIERS APPEARING IN STARTUPS CAUSING SLOWER INNOVATION SPEED

| Startup | Barriers | | | | | |
|---|---|---|---|---|---|---|
| | Late involvement of software developers | Executive sponsor is missing or not clarified | Yearly budgeting and planning | Unclear decision-making authority | Lack of digital infrastructure for experimentation | Access to data from external actors |
| Startup 1 | x | x | | x | x | x |
| Startup 2 | x | | x | x | x | |
| Startup 3 | x | x | | x | x | |
| Startup 4 | (x) | x | | x | x | |
| Startup 5 | x | (x) | (x) | (x) | x | |

*Note.* x = barrier registered; (x) = barrier registered, but was addressed during the data-collection period

First, not identifying and mitigating technical risks early, forced product managers in startups 2 and 5 to redesign their solutions because developers identified them as impossible to develop later and integrate with the existing system. We found that the product managers in charge of the novel product ideas did not focus on the technical aspects in the early phases. Their approach was to validate the idea by extensive customer interviews and by presenting mockups and PowerPoint presentations in Microsoft Teams meetings. While the product managers got a lot of good customer insight, they were often uncertain if their idea was technically feasible. One product manager expressed: *"It's all about bridging a need in the market to a piece of working software, but how do I determine if it's possible to make?"* Consequently, product managers invested in developing ideas that were later found not technically feasible or so complicated that it would require a long time to develop. Long development times could result in another competitor picking up the idea as the customer in the global market frequently communicated with MarComp's competitors. Consequently, the product managers expressed a need for earlier collaboration with software developers to identify technical issues before wasting time on a non-feasible idea.

The second challenge with a disconnected process between business and development appeared after the product managers had verified the business potential and were ready to start building the new solution. While they had a vague idea about what software developers needed to build the solution, this was not enough to achieve progress. For a software team to implement the idea, they needed concrete user-stories, prioritized feature-lists, security analysis, etc. As this was not thought of, many preparations and discussions had to be conducted before the actual development activities could start. One product manager explained: *"if IT had been there from the start asking relevant questions […] I would know what to deliver and kept continuity."* Another product manager explained that continuous dialogue with a software developer would guide him towards knowing what was expected of him by the software developers. Often, product managers had gathered much of the necessary customer insight. However, it became evident that much effort was needed to transfer the insight to concrete features and user stories that the developers could work on.

Further, there was a need to analyze such features in detail to understand the development effort needed. Without such knowledge, it became hard to prioritize features and to build minimum viable products (MVPs). One product manager explained, *"You get a lot of insights into the customer's needs [in the phases before Build], but I found it difficult to translate it to functional requirements."* Consequently, when entering the Build-phase, the innovation process was hampered because of the need to onboard software developers and prepare for software development. The delay resulted in slower progress and consequently reduced motivation of the product managers. One product manager confessed: *"It kills my motivation when we have to redo what we did months ago prior to onboarding our new participants [software developers]. We have used four weeks with no actual progress, I know we could've seriously shortened onboarding time if just some of these developers participated some months ago."*

One mitigating strategy against the late involvement of software development among the product managers was to use their personal network to invite software developers they already knew to assist them. The product manager in startup 4 said: *"I have a programming background myself, so I just used my network and invited someone I knew, and he drew on the resources he needed."* At the same time, product managers with limited networks needed assistance in finding available developers.

*B. The executive sponsor role is missing or not clarified*

To progress, the startups needed support from executive sponsors to remove organizational barriers, such as disagreements among line managers in a different department, and limited authority to make business and financial decisions. Startups were sometimes in conflict with existing products or needed many departments to collaborate. Cooperation across departments was cumbersome because of the need to involve many people in decision-making, e.g., agreeing on using resources or how to scale a startup. As some departments were spread worldwide and even had different key performance indicators (KPIs), the decision-making process became even more cumbersome. It was clear that a sponsor role was needed to ensure smoother decision-making per startup. Additionally, executive sponsors often had an extensive network within the company, which could help product managers to find needed competence from other departments. Despite the encouragement in the innovation framework to assign a dedicated executive sponsor to each startup, the role was often missing, and the responsibilities of the sponsor role were unclear to the product manager.

This issue was successfully addressed in startup 5, where the sponsor decided to fund software development resources over the sponsor's budget. Emerging funding was beneficial because it reduced waiting time. The IT resources were usually allocated through yearly budgeting (see chapter "Yearly budgeting and planning" below). Therefore, the sponsor's funding provided the startup with access to software development in a more flexible and timely way. While such a decision speeded up the specific startup, the product manager remained unaware of how the sponsor suddenly could get additional funding. She stated: "*Before Summer, I started to figure out what resources we needed [for the software development], but why it did not happen then, but happened now, that I do not know*." Not knowing that the sponsor could help secure more money indicated a lack of role clarity and collaboration between the product manager and the sponsor. The product manager was not aware that the sponsor could bypass the existing budgeting processes.

*C. Yearly budgeting and planning*

The practice of yearly budgeting and planning periods were barriers to innovation in two ways. First, innovations moving into Build-phase (which could happen anytime during the year) had to compete for budget with other ongoing department-activities, risking long waiting times before being prioritized into the units' portfolio. A startup was initially funded by the department where the product manager was situated, then transferred to an operation department when entering the Build-phase. Before the Build-phase, startup-participants are expected to utilize flexibility within their own schedule, meaning that their department allows them to use internal time, effectively funding the startup. The idea of handing the startup over to an operational department meant that any specific business department would not fund a startup before starting software development activities. However, since software development required significant funding, there could be long waiting times for some startups as ongoing projects needed to be reprioritized. Moreover, the ongoing projects were often critical and could not be delayed, or they were already late. Further, existing budgets and plans in each department were formulated annually. Therefore, when a new startup entered the department, existing decisions had to be redone.

As mentioned earlier, startup 5 bypassed the budgeting process and got extra funding from the executive sponsor. However, allocating more money to the startup gave rise to another challenge; software developers did not have capacity for an extra project. The architect assigned to the startup described the challenge: *"More funding doesn't give me two more hands to work with, I still don't have time for this."* Because of long-term plans being laid out, developers were fully booked for an equally long period, and lacked flexibility to prioritize the emerging innovations.

Second, key experts needed for the innovation process were usually fully booked on several ongoing projects and therefore lacked the flexibility to respond to product managers' requests promptly. We saw that key personnel from other departments (e.g., finance, IT-infrastructure, business development, and marketing) who were requested to support startups did not have sufficient flexibility and could not respond quickly to the need of a startup. Not getting access to experts resulted in delays in the innovation work, as the product managers had to either wait in line for the key personnel or find workarounds.

*D. Unclear decision-making authority*

Even though product managers were the main drivers behind the startups, it was unclear how much actual decision-making authority they held. The unclear authority of the product managers was a barrier, given that the progress of the startups depended heavily on many strategic, tactical, and operational decisions, such as access to development resources. Further, strategic (goals and objectives) and tactical decisions (resource allocation) needed to be taken frequently, while operational decisions (solving tasks) needed to be taken daily. Many line managers wanted to be involved in decision making, so the product managers spent much time on anchoring decisions among the line managers (such as marketing, finance, business development), which was a complicated process. One product manager explained: "*It is sometimes challenging because there are so many people that have opinions on what we are doing*."

Further, the product managers worked only part-time on their startup, which increased the decision-making problem because it took a long time to align key stakeholders. Besides, product managers received divergent inputs from line managers. In one example, the line managers and sales experts had different opinions on which pricing model the startups should opt for. Such disagreement led to delays in the innovation process since the product managers needed to validate the decisions with numerous stakeholders instead of working on the startup. To speed up the decision-making process, the product manager in startup 5 established a steering committee where key stakeholders advised her, took critical strategic and tactical decisions, and made sure that the decisions aligned with the company's various business objectives. The product manager explained, *"[…] it feels safer to take decisions as a committee because I feel that I have a backing to move forward."* Further, she explained that when line managers sought to influence or challenge a decision, she referred them to the steering committee instead of handling them herself. This showed to be a successful strategy securing her time to focus on the innovation.

*E. Lack of digital infrastructure for experimentation*

There was a need for a digital platform that enabled experimenting with startups. A prerequisite for continuous experimentation is to have the right resources such as data, development resources, and an infrastructure that enables fast experimentation. Product managers ran numerous customer interviews and experiments (e.g., presentations of mock-ups) to figure out the value of the startups to potential customers. However, the product managers were working independently from each other, which resulted in much time wasted for producing the experimentation materials (e.g., UX-design, interview guides, etc.). One product manager stated, "*some of the software we developed for the experiment is custom made and is not reusable for others, but the work done by UX-designers are reusable and can save time for other product*

*managers."* Many resources could be saved if different product managers had access to a shared pool of such materials.

Furthermore, the product managers experimented with pricing models to determine how much the customers were willing to spend to access the potential services. This was challenging because the platform functionality needed for such experiments was lacking (e.g., current payment methods provided by the customer portal were limited). As it became evident from a Community of Practice-meeting, only two of the five available pricing models could be implemented in the customer portal today.

*F. Access to data from external actors*

All five startups relied on accessing data from ships and subcontractors of ship equipment. Challenges in accessing data became a considerable barrier for Startup 1. The new business idea needed data that was automatically measured and stored in an installed device on ships. Gaining access to the data was possible only manually (e.g., through USB-stick into the device and downloading), because the vendor of the data-producing device was reluctant to share the data. The product manager said: "*Shipowners are more than willing to share their data, while the device vendor is very much aware of the data's value.*" The data was owned by the shipowner who operated the device, but only until it was uploaded to the vendor's cloud-service; then the vendor claimed ownership. The product manager had several meetings with the vendor to secure collaboration and set up an experiment on extracting data from the cloud service. At the beginning of the collaboration, the vendor was enthusiastic and viewed the collaboration as a win-win project. However, later the vendor started asking questions about the business case and became reluctant to share data, supposedly becoming aware that the data had high value. The collaboration was put on hold and resulted in a lengthy negotiation process between MarComp's and the vendor's lawyers, which is still ongoing. Since then, there has been no progress in the startup for several months.

V. DISCUSSION

Large global companies need to speed up their innovation activities to increase competitive advantage. One suggested way to speed up software product development is to continuously integrate business and development [12]. The results of this study seek to shed light on the research question: *What are the barriers to software product innovation speed in a large company?* We have presented our findings on five internal startups in a global provider of maritime services, where six main barriers to software product innovation are identified: *late involvement of software developers*, *executive sponsor is missing or not clarified*, *yearly budgeting and planning*, *unclear decision-making authority*, *lack of infrastructure for experimentation*, and *access to data from external actors*. Together, these barriers appeared to slow down the innovation speed within the examined startups.

To answer the research questions, we will now discuss the barriers, their meaning for the innovation speed, and how they relate to the existing literature. Further, we will examine our findings through the lens of continuous software engineering and especially the integration between development and business strategy (BizDev) [12]. We end the discussion by explaining how a continuous software engineering perspective can help GSE companies increase innovation speed.

*A. Barriers and their meaning for the software product innovation speed*

As shown in table II some barriers were identified in all startups, while some only appeared in a few. Late involvement of software developers and unclear decision-making authority was a common barrier, while access to data from external actors (shipowners and vendors) only hindered one startup. Despite sharing a common context (the maritime sector) and the same innovation framework, the startups still faced different barriers, showing there is no clear-cut process for conducting software product innovation in globally distributed companies. The startups acknowledged the barriers to lower their innovation speed and tried to find ways of overcoming them. Despite often facing the same barriers, they handled them differently. Some experienced product managers used their network and personal connections to secure the early involvement of software developers.

In contrast, less experienced product managers relied on their executive sponsors' network to come in contact with developers and other external specialists. Another example was establishing a committee of stakeholders to align decisions between different units to address the barrier of unclear decision-making authority. The other startups toiled with this barrier, using much time anchoring their decision with line managers. Further, overcoming one barrier where shown to introduce another. For example, funding a startup outside the normal budgeting cycle gave rise to the barrier of infrequent planning as a software developer became overbooked. Finally, all barriers appeared to slow down the innovation speed. The following paragraphs elaborate on the barriers, and we compare them to other relevant findings in the literature.

Large global companies tend to be complex and bureaucratic, leading the software product innovation to slow down due to several layers of approval [3]. For example, management factors (e.g., corporate guidelines, potential internal conflicts in the company) have been shown to slow down internal startups' development process [15]. In our case, management factors were also reducing the innovation speed. For example, the lack of explicitly identified decision-makers led to repeated delays in the startup development because product managers needed to consider the viewpoints of a wide variety of stakeholders. Some of them had conflicting views on the desirability of the startups' features and functionality. Our findings are thus in line with previous results by Moe, Aurum & Dybå [28], that found that aligning strategic, tactical, and operational decisions was complicated when the stakeholders were not aligned. We conclude that such alignment is crucial for the speed of software product innovation in global companies.

Our findings draw particular attention to the business-software integration in the context of software product innovation and are thus in line with previous literature. The disconnect between business strategy and software development has been frequently discussed in software engineering [12], [29]. Previous research has shown that software developers understand the need to be involved early and desire to get involved in strategic business decision-making rather than consulted after decisions have been taken [28]. However, such involvement is often missing in practice [12]. Software developers in our case study were not involved in the majority of startups before the Build-phase, which led to significant delays in the innovation process, as product managers needed to revisit technical requirements and UX-design. This is in line with previous findings that continuous access to expertise from other units is a challenge in large companies [25]. To address this challenge, people usually rely on their personal networks [25], [26], which we also found in one startup. Nevertheless, late involvement of development was generally a problem in other startups, which was especially crucial in the innovation context since it appeared to increase the products' time to market. We thus speculate that delays in software product innovation can be caused by weak interaction between business and development at early stages.

Further, our results highlight the executive sponsorship's role in the internal innovation process (e.g., Lean startup). The majority of the described startups lacked a clear executive sponsor, which appeared to slow down the decision-making process. At the same time, it was beneficial for the speed of one startup when the sponsor was active (e.g., startups gaining access to software development resources). Executive sponsorship or championship appears crucial for innovation in large companies [30] and is known to be enabling Lean internal startups as they protect them from the negative influence of corporate policies and changes in strategy [15]. Startups without influential champions lack the necessary support and may run the risk of being terminated [5]. Influential executive sponsors have access to broad social networks and can thus provide startup teams with access to necessary expertise, which is often impaired in large companies [24]–[26]. This led us to suggest that active executive sponsors in large and global companies play an essential role in software product innovation and its speed.

To build and conduct meaningful experimentation (the 4$^{th}$ principle of lean Startup), close collaboration between business and software development is needed. Implementation of the build-measure-learn loop in large companies can be challenging because it brings the innovating teams to conflict with its ongoing procedures [5]. An experimentation system requires the ability to release minimum viable products or features, design and manage experiment plans, link experiment results with a product roadmap, and manage a flexible business strategy [14]. Our data are in line with these findings and show that continuous experimentation is dependent on rapid and proper design of experiments. Personnel from both business and software development worked together to build technical components and business hypotheses from scratch every time, which demanded a lot of time and resources. Lack of shared infrastructure for experimentation can thus reduce the innovation speed in large companies.

Finally, we highlight that access to external data can be a strong but hidden impediment to software product innovation. There is little empirical research into data partnerships and how business relations are handled in data-sourcing today. What little exists report that highly standard "as is" contracts or informal tacit contracts are used [31]. In our case vendor's unwillingness to share data from vessels' data-recording device became a crucial barrier that put the startup development on hold for several months. We emphasize that access to data moves away from being a technically related barrier and towards a business issue that might lead to lengthy legal negotiations between numerous actors globally. We thus point out that access to data from external actors is an essential aspect of software product innovation in GSE companies.

We have shown that the barriers are slowing innovation speed. Some barriers are well known and described (e.g., yearly budgeting and planning, late involvement of software development), while others are not often emphasized in the innovation context (e.g., the barrier of accessing data). We will now turn to the discussion of our findings through the lens of the conceptual framework.

*B. Continuous software engineering in software product innovation: the role of BizDev and the innovation speed*

The findings that late involvement of software development in working with new business ideas impairs the innovation process relates to poor integration between Business Strategy and Development as described in the continuous software engineering framework [12]. The authors argue that such integration (BizDev) is a prerequisite for software development flow and that the BizDev is a common challenge in software firms. Our findings support these claims in the GSE context, as they illustrate how the stakeholders involved in the initial work with the startups (Business Strategy) lacked input from the software unit (Development). For example, potential software development resources were not taken into account when planning the yearly budget. Further, the software developers' perspective was not considered during the early phases of the innovation work (e.g., concept validation with potential customers). Such disintegration was challenging for the internal startups, highlighting the importance of early BizDev integration, especially in software product innovation.

The shortage of development resources in the company, and infrequent budgeting often put the startups on hold until new plans and budgets were laid. The authors of the framework argue that planning and budgeting in software firms need to be continuous to ensure a better development flow. Our findings support this reasoning, showing that episodic budgeting leads to slower innovation speed. There is a general view that speed is crucial for product innovation [19]. However, the authors of the framework show that inadequate up-front planning of the innovation process is a significant cause of failure for companies delivering customer products [12], making continuous planning and budgeting a prerequisite for software product innovation.

The startups in our study had various speeds at different times. Periods of intense activity were followed by periods of inactivity due to the periodically encountered organizational barriers. The authors of the continuous software engineering framework argue that there is a misplaced focus on speed rather than continuity in software product development [12]. Through our case, we have observed that slow time to market had to do with a lack of continuity rather than low speed and that organizational barriers caused the lack of continuity. For example, the product managers often had to wait for evaluation from several line managers to proceed in their innovation work. This leads us to emphasize the role of continuity in the software product innovation process. An often-cited metaphor is the story of the slow but consistent tortoise that wins the race against the speedy but occasionally dozing hare. In line with continuous software engineering principles, we find that the speed alone is insufficient to reduce time to market unless continuity is also in place.

*C. Implications for practice*

In this work, we addressed the challenge of bringing new products to the market faster in large distributed companies. By arguing that continuity deserves significant focus when trying to increase innovation speed, we recommend companies to strengthen the integration between business activities and software development activities.

As a recommendation for practitioners, we present measures that Marcomp's decided to implement to improve innovation speed while trying to avoid coming in conflict with operations:

- Map internal barriers and challenges that reduce innovation speed to implement measures. Use, e.g., Table II and the list of barriers we found in this work.
- Involve software developers early to support product managers by providing feedback on the idea and help them understand what is doable if the idea enters the scale-up phase. Developers should then primarily play an adviser's role and not engage in development activity this early to spare resources. This recommendation is in line with Lehtola et al. [29] that suggest greater collaboration between developers and other environments, such as business, sales and marketing and R&D, earlier in development cycles.
- Increase the startup's authority. When decision-making is decentralized and teams becomes more autonomous, it will increase speed as decisions can be taken at late stages by those doing the innovative work [28].
- Build a self-service platform where startups can set up experiments by themselves. Speed will probably increase when startups can efficiently experiment without being dependent on others [32].
- Convert to continuously plan and budget to make room for emerging startups and try Beyond budgeting principles [33].
- If startups are dependent on external stakeholders' data, we propose increased attention towards accessibility early to avoid time-wasting while developing a product.

We believe these recommendations will assist practitioners in innovating even faster within a global software-intensive company.

## VI. CONCLUSION AND FUTURE WORK

We have presented findings on five internal startups' innovation processes to determine which barriers reduce software product innovation speed in large global companies. Innovation speed is crucial for global software-intensive companies to stay competitive. Our findings demonstrate how innovation processes slow down due to the surrounding organizational context and that maintaining continuity is a crucial factor for innovation speed. We have concluded that early integration between business and software development activities (known as BizDev) contributes to innovation speed and continuity. Our findings can assist organizations in planning and improving their software product innovation activities.

This study contributes to the literature by showing that the framework on continuous software engineering was well-functioning in making sense of our findings. It supported us in finding barriers to innovation speed by searching for breaks in continuity. Also, the study contributes by providing empirical evidence on implementation of the Lean startup approach in large organizations, called for by Edison et al. [15].

Future research should keep investigating barriers to software product innovation to widen our understanding of the process and assist practitioners in their innovation work. Specifically, we urge researchers to study how accessibility to necessary data from external actors gives rise to new business-related issues like data partnership. Also, we call for empirical investigations into self-service platforms as infrastructure for continuous experimentation. Findings on later stages of the case startups will follow as the startups proceed further in their innovation process and establish products in the market.